# Approximation Algorithms for the Partition Vertex Cover Problem

Suman K. Bera[1], Shalmoli Gupta[2], Amit Kumar[3], and Sambuddha Roy[1]

[1] IBM India Research Lab, New Delhi
[2] University of Illinois at Urbana-Champaign
[3] Indian Institute of Technology Delhi

**Abstract.** We consider a natural generalization of the `Partial Vertex Cover` problem. Here an instance consists of a graph $G = (V, E)$, a cost function $c : V \to \mathbb{Z}^+$, a partition $P_1, \ldots, P_r$ of the edge set $E$, and a parameter $k_i$ for each partition $P_i$. The goal is to find a minimum cost set of vertices which cover at least $k_i$ edges from the partition $P_i$. We call this the `Partition-VC` problem. In this paper, we give matching upper and lower bound on the approximability of this problem. Our algorithm is based on a novel LP relaxation for this problem. This LP relaxation is obtained by adding knapsack cover inequalities to a natural LP relaxation of the problem. We show that this LP has integrality gap of $O(\log r)$, where $r$ is the number of sets in the partition of the edge set. We also extend our result to more general settings.

## 1 Introduction

The `Vertex Cover` problem is one of the most fundamental NP-hard problems and has been widely studied in the context of approximation algorithms [21, 10]. In this problem, we are given an undirected graph $G = (V, E)$ and a cost function $c : V \to \mathbb{Z}^+$. The goal is to find a minimum cost set of vertices which *cover* all the edges in $E$ : a set of vertices $S$ covers an edge $e$ if $S$ contains at least one of the end-points of $e$. Several 2-approximation algorithms are known for this problem [1, 16]. The `Partial Vertex Cover` problem is a generalization of the `Vertex Cover` problem, where we are also given a parameter $k$. The goal is to find a minimum cost set of vertices which cover at least $k$ edges. This problem was proposed by Bshouty and Burroughs [4], and they gave a 2-approximation for this problem using LP-rounding. Since then, many different techniques have been shown to give 2-approximation algorithm for this problem ( [17, 2, 13]).

In this paper, we consider a natural generalization of the `Partial Vertex Cover` problem. Here an instance consists of a graph $G = (V, E)$, a cost function $c : V \to \mathbb{Z}^+$, a partition $P_1, \ldots, P_r$ of the edge set $E$, and a parameter $k_i$ for each partition $P_i$. The goal is to find a minimum cost set of vertices which cover at least $k_i$ edges from the partition $P_i$. We call this the `Partition-VC` problem. In this paper, we give matching upper and lower bound on the approximability of this problem. We give an $O(\log r)$-approximation algorithm for the `Partition-VC` problem, and show that unless P=NP, we cannot do better. Recall that $r$ denotes the number of sets in the partition of the edge set $E$. Note that for constant number of partitions this gives a constant approximation. Our results also extend to the slightly more general problem where edges have weights, and we would like to pick a minimum cost set of vertices which cover edges of total weight at least $\Pi_i$ for each partition $P_i$. We call this the `Knapsack Partition Vertex Cover` problem.

**Our Techniques:** The hardness result for the `Partition-VC` problem follows by an approximation preserving reduction from the `Set Cover` problem. The approximation algorithm uses a novel LP relaxation – this is the main contribution of the paper, and we expect this idea to have more applications. The natural LP relaxation for even the `Partial Vertex Cover` turns out to be unbounded. Indeed, consider the following example : the graph is a star – there is a vertex $v$ of degree $D$ and all its neighbors are leaves. All vertex costs are 1, and the parameter $k = 1$. Clearly, any optimal solution must cost at least 1 unit. But a fractional solution will pick the vertex $v$ to an extent of $\frac{1}{D}$, and hence will cover all the $D$ edges fractionally to an extent of $\frac{1}{D}$. So, the fractional solution pays only $\frac{1}{D}$. One way of getting around this problem is to augment the LP with more information. Here, we can guess the most expensive vertex an optimal solution will pick, and can remove all vertices with cost more than the cost of this vertex. Further, the cost of this vertex is also a lower bound on the optimal value. This idea was used by [13] to give a 2-approximation for the `Partial Vertex Cover` problem. However, applying such an idea to the `Partition-VC` problem turns out to be non-trivial. We cannot guess the most expensive vertex in each of the partitions – this will take time exponential in $r$. Our approach is to strengthen the natural LP relaxation such that no guesswork is required. We show how to do this using knapsack cover inequalities [7]. Armed with this stronger relaxation, we show that one can carefully use randomized rounding based techniques to get the approximation algorithm.

**Related Work:** There has been recent work on partial covering versions of several covering problems. For the set cover problem, the partial covering version namely the partial set cover problem was first studied by Kearns [18], who proved that the approximation ratio of the greedy algorithm is at most $2H(n)+3$, where $n$ is the size of the ground set and $H(n)$ is the $n^{th}$ harmonic number. Later Slavík [20] showed that it is actually bounded by $H(k)$ where $k$ is the number of elements to be covered. This natural greedy approach when extended for the `Partition-VC` problem gives only a $H(|V|)$-approximation, which is much worse than the lower bound of $O(\log r)$ that we have proved for this problem.

The partial vertex cover problem has also been widely studied in the literature. Bshouty and Burroughs [4] were the first to give a polynomial time 2-approximation algorithm for it. Subsequently, several other algorithms based on Lagrangian Relaxation, local-ratio, primal-dual techniques with the same approximation guarantee were proposed [17, 2, 13, 19]. Mestre's [19] primal-dual technique can also be used to get a 2-approximation for a more general version of the problem, the partial capacitated vertex cover problem. Bar-Yehuda et al. [3] gave constant factor approximation algorithms for several variants of this prob-lem using the local-ratio technique. Partial versions have also been studied for the Facility Location problem and its variants. Charikar et al. [8] explored the outlier or robust version of the uncapacitated facility location problem and k-center problem where only a fraction of the clients need to be serviced. For both the problems they gave constant approximation algorithms. Apart from these there are several other partial covering problems: e.g. k-median with outliers [9], k-MST problem [14] and k-multicut problem [15]. However, these approaches do not seem to work for the `Partition-VC` problem.

The set of partitions of the edge set in the `Partition-VC` problem are a special case of matroids. There has been significant work on maximizing a submodular function under matroid constraints [22, 5], but none of these results apply to the `Partition-VC` problem.

The rest of the paper is organized as follows. We present the hardness of the `Partition-VC` problem in Section 2. Our rounding algorithm and its analysis is presented in Section 3. Finally concluding remarks are made in Section 4.

## 2 Hardness of the `Partition-VC` problem

In this section, we prove that it is NP-hard to get better than $O(\log r)$-approximation for the `Partition-VC` problem. We give a reduction from the `Set Cover` problem. Recall that an instance $\mathcal{I}$ of the set cover problem consists of a set $X$ containing $r$ elements, and a set $\mathcal{S}$ of subsets $S_1, \ldots, S_m$ of $X$. The goal is to find minimum number of sets in $\mathcal{S}$ such that their union is $X$. This problem is known to be NP-hard, and in fact, it is known that unless P=NP, any polynomial time algorithm for the `Set Cover` problem must have approximation ratio of $\Omega(\log r)$ [12].

We now describe the reduction in detail. Let $\mathcal{I}$ be an instance of the `Set Cover` problem as described above. We now construct an instance $\mathcal{I}'$ of the `Partition-VC` problem. The graph $G'$ in $\mathcal{I}'$ is a bipartite graph. The vertices on left side, $V'_L$ are defined as follows : for each set $S_i \in \mathcal{S}$, we add a vertex $s'_i$ to $V'_L$. All these vertices have unit cost. The vertices on the right side, $V'_R$ are as follows : for every element $u \in X$, we have a corresponding vertex $u' \in V'_R$. Each of these vertices has infinite cost. Now, we define the set of edges $E'$ and the partition $P_1, \ldots, P_r$ (note that the number of sets in the partition is same as the size of the set $X$ in $\mathcal{I}$). For a vertex $s'_i \in V'_L$ and $u' \in V'_R$, we have an edge between them in $E'$ iff $u \in S_i$ in the instance $\mathcal{I}$. We partition the set $E'$ as follows : for every $u' \in V'_R$, define $P_{u'}$ as the set of edges incident to $u'$. The partition of $E'$ is $\{P_{u'} : u' \in V'_R\}$. Further, the quantities $k_{u'}$, which tell how many of the edges in the set $P_{u'}$ need to be covered, are 1. This completes the description of the instance $\mathcal{I}'$. The following lemma is now easy to see.

**Lemma 1.** *There is a solution to $\mathcal{I}$ of cost $C$ iff there is a solution to $\mathcal{I}'$ of cost $C$. Hence, unless P=NP, any polynomial time algorithm for the `Partition-VC` problem must have approximation ratio of $\Omega(\log r)$.*

*Proof.* Any solution to $\mathcal{I}$ picks a some subsets $S_{i_1}, \ldots, S_{i_l}$ in $\mathcal{S}$. Then we can have a solution of the same cost for $\mathcal{I}'$ in which we pick the corresponding vertices in $V'_L$. Similarly, consider a solution to $\mathcal{I}'$. None of the vertices picked by this solution can be in $V'_R$ (because of infinite cost). Thus, we can look at the corresponding subsets in $\mathcal{S}$, and these subsets will form a set cover in $\mathcal{I}$.

## 3 Approximation Algorithm for the `Partition-VC` problem

In this section, we give an $O(\log r)$-approximation algorithm for this problem. We begin with the natural LP relaxation, and then strengthen it by adding knapsack cover inequalities. Fix an instance $\mathcal{I}$ consisting of a graph $G = (V, E)$, partition $P_1, \ldots, P_r$ of $E$, and parameters $k_1, \ldots, k_r$. The natural LP relaxation is described below. For every vertex $v \in V$, we have a variable $x_v$ which should be 1 if we pick this vertex, and 0 otherwise. For an edge $e$, we have a variable $y_e$, which should be 1 if $e$ gets covered by the solution, 0 otherwise.

$$\min \sum_{v \in V} c_v x_v \qquad \text{(LP1)}$$
$$x_u + x_v \geq y_e \qquad \text{for all edges } e = (u,v) \in E$$
$$\sum_{e \in P_i} y_e \geq k_i \qquad \text{for every partition } P_i$$
$$0 \leq x_v \leq 1 \qquad \text{for all vertices } v \in V$$
$$0 \leq y_e \leq 1 \qquad \text{for all edges } e \in E$$

As explained in the introduction, the integrality gap of this LP relaxation is unbounded even for the `Partial Vertex Cover` problem.

To improve the integrality gap, we add knapsack cover inequalities as follows. Consider a subset of vertices $A$. Suppose we select all the vertices in $A$. Now, $A$ will cover some of the edges in each of the partitions $P_i$. Define $k_i(A)$ as the number of edges we still need to cover from $P_i$ (after having picked $A$). So, we must choose enough vertices from $V \setminus A$ such that the remaining covering constraint in every partition is met. In other words, the following conditions must be satisfied

$$x_u + x_v \geq y_e \qquad \forall e = (u,v) \in E \text{ and } u, v \notin A$$
$$\sum_{\substack{e=(u,v) \in P_i \\ u,v \notin A}} y_e \geq k_i(A) \qquad \text{for every partition } P_i$$

If we replace the variable $y_e$ in the second inequality above by using the first inequality, we get that for every partition $P_i$,

$$\sum_{\substack{e=(u,v)\in P_i \\ u,v\notin A}} (x_u + x_v) \geq k_i(A)$$

$$\text{i.e., } \sum_{v\notin A} \deg_i(v, A) x_v \geq k_i(A),$$

where $\deg_i(v, A)$ denotes the degree of $v$ in the subgraph of $G$ considering edges in $P_i$ only and removing the vertices of $A$. Since $x_v$ have value lying the range $[0, 1]$, we can further strengthen the above by truncating the values $\deg_i(v, A)$ to $\min(\deg_i(v, A), k_i(A))$. Thus we get the following strengthened LP relaxation:

$$\min \sum_{v\in V} c_v x_v \qquad\qquad \text{(PVC-LP)}$$

$$\sum_{v\notin A} \min(k_i(A), \deg_i(v, A)) x_v \geq k_i(A) \quad \text{for all partitions } P_i, \text{ subsets } A \subseteq V \tag{1}$$

$$0 \leq x_v \leq 1 \quad \text{for all vertices } v \in V$$

We shall show that the integrality gap of the above LP relaxation is $O(\log r)$. We first present the rounding algorithm, and then we will discuss how to solve this LP. So assume that we have a solution $x^\star$ to the LP above. The rounding algorithm is described below.

---

**Algorithm 1:** Rounding a solution to the PVC-LP

Given a solution $x^\star$.
Let $\hat{x}$ be the integral solution that we will build. Initially, $\hat{x}_v = 0$ for all $v$.
**for** $\forall v \in V$ **do**
  **if** $x_v^\star \geq 1/6$ **then**
    Set $\hat{x}_v \leftarrow 1$
  **else**
    Set $\hat{x}_v \leftarrow 1$ with probability $6x_v^\star$.
  **end if**
**end for**
Pick the set of vertices $v$ in the solution for which $\hat{x}_v = 1$

---

**Analysis of the Rounding Algorithm**

We begin by showing that the solution constructed above is good for any partition $P_i$ with constant probability.

**Theorem 1.** *For any partition $P_i$, the solution $\hat{x}$ covers at least $k_i$ edges of $P_i$ with probability at least $5/8$.*

*Proof.* Define $A$ to be the set $\{v \in V : x_v^* \geq 1/6\}$. Note that our algorithm picks all the vertices in $A$. Therefore, we just need to show that the vertices picked from $V \setminus A$ cover at least $k_i(A)$ edges from $P_i$ after we remove the vertices in $A$. Let $\beta_i(v, A)$ denote $\frac{\min(\deg_i(v, A), k_i(A))}{k_i(A)}$. Then the constraint (1) applied to this particular set $A$ implies that

$$\text{(2)} \qquad \sum_{v \notin A} \beta_i(v, A) x_v^* \geq 1.$$

**Lemma 2.** *For any partition $P_i$,*

$$\sum_{v \notin A} \beta_i(v, A) \hat{x}_v < 2,$$

*happens with probability at most $3/8$.*

*Proof.* The proof is a simple application of Chebychev's inequality and uses the fact that the quantities $\beta_i(v, A)$ are at most 1. For a vertex $v \notin A$, let $Y_v$ be an indicator random variable which is 1 if $v$ is included in the solution (i.e., $\hat{x}_v = 1$), 0 otherwise. Let $Z_i$ denote $\sum_{v \notin A} \beta_i(v, A) Y_v$. The expected value $E[Z_i]$ can be expressed as

$$E[Z_i] = \sum_{v \notin A} 6 \beta_i(v, A) x_v^\star \geq 6,$$

using the inequality (2). We now bound the variance $Var(Z_i)$ of $Z_i$.

$$Var[Z_i] = \sum_{v \notin A} \beta_i(v, A)^2 Var(Y_v)$$
$$= \sum_{v \notin A} \beta_i(v, A)^2 \cdot 6 x_v^\star (1 - 6 x_v^\star)$$
$$\leq 6 \sum_{v \notin A} \beta_i(v, A) x_v^\star \quad \text{because } \beta_i(v, A) \leq 1$$

The claim now follows from Chebychev's inequality. Indeed, we want to bound the probability $\Pr[Z_i < 2]$. This can be done as follows :

$$\Pr[Z_i < 2] \leq \Pr[Z_i < E[Z_i]/3] \leq \Pr[|Z_i - E[Z_i]| \geq \frac{2 E[Z_i]}{3}]$$
$$\leq \frac{9}{4} \cdot \frac{Var[Z_i]}{E[Z_i]^2}$$
$$\leq \frac{9}{4} \cdot \frac{6 \sum_{v \notin A} \beta_i(v, A) x_v^\star}{\left(\sum_{v \notin A} 6 \beta_i(v, A) x_v^\star\right)^2} \leq \frac{3}{8},$$

where the last inequality uses (2).

Now, suppose the solution $\hat{x}$ does not cover at least $k_i$ edges of $P_i$. Then, restricted to the sub-graph of $G$ where we include edges in $P_i$ only, and remove all vertices in $A$, the total degree of the vertices picked by our algorithm (in $V \setminus A$) will be less than $2k_i(A)$. In other words,

$$\sum_{v \notin A} \beta_i(v, A)\hat{x_v} < 2.$$

But lemma 2 shows that the probability of this event is at most $3/8$. This proves the theorem.

Thus, in expectation, more than half of the partitions get satisfied. To satisfy all the partitions, we just repeat our algorithm $O(\log r)$ times. So, our final algorithm is : repeat Algorithm 1 $c \log r$ times, where $c$ is a large constant. We output the union of all the vertices chosen in each such round. The following theorem now shows that our algorithm is an $O(\log r)$-approximation algorithm.

**Theorem 2.** *With high probability, the algorithm outputs a feasible solution and its cost is $O(\log r) \cdot \sum_{v \in V} c_v x_v^\star$.*

*Proof.* Lemma 2 shows that in any particular round, we cover at least $k_i$ edges of $P_i$ with probability at least $5/8$. So, the probability that we do not satisfy the constraint for $P_i$ in any of the rounds is at most $1/r^{c'}$ for some large constant $c'$, and hence, by union bound, our algorithm outputs a feasible solution with high probability. Also, the expected cost of each round is at most $6 \sum_{v \in V} c_v x_v^\star$. This proves the theorem.

**Solving the LP Relaxation** Finally, we show how we can get a solution $x^\star$ for the PVC-LP. We first guess the value of the optimal solution – call it $\Delta$ (we can always do this up to any constant precision by binary search). We convert the LP to a feasibility LP by removing the objective function, and adding a constraint

$$\sum_v c_v x_v \leq \Delta.$$

Now, we use the ellipsoid method to solve the LP. Given a candidate solution $x$, we first check if it satisfies the above constraint – if not, we can just return this violated constraint. Otherwise, we define $A = \{v : x_v \geq 1/6\}$. We check the constraint (1) for this set $A$, and again, if this is not satisfied, we can return this as a violated constraint. Now, notice that our rounding algorithm just requires the solution $x^\star$ to satisfy these two inequalities, and we need not even check all the (exponentially many) constraints (1).

### 3.1 Extensions

We now show that our result can be extended to more general settings.
**The Knapsack Partition Vertex Cover problem :** Recall that in this problem, we have weights $w_e$ associated with each edge $e$. Again given a partition $P_1, \ldots, P_r$, and parameters $\Pi_i$, we would like to pick a minimum cost subset of vertices such that they cover edges of cost at least $\Pi_i$ from the set $P_i$ for each

$i$. Our algorithm and analysis extend in straightforward way to this setting as well.

**The sets $P_i$ need not be disjoint :** Our analysis does not require these sets to be disjoint. The same algorithm works here as well. Note that our hardness results holds in the stronger setting where we want these sets to be disjoint.

## 4  Conclusion

We have presented algorithms for the `Partition-VC` problem. For this problem using primal-dual schema similar to the one described by Tim Carnes & David Shmoys [6] we can obtain an $O(f)$-approximation algorithm, where $f$ is the maximum number of edges in a partition $P_i$. The proof is quite straight forward. This result is analogous to the $f$-approximation result for the `Set Cover` problem [11, 1]. It will be interesting to extend our techniques to partition versions of other partial covering problems. One natural related problem is the `Partition Set Cover` problem. The `Partition Set Cover` problem can be seen as a generalization of the `Partial Set Cover` problem where $P_1, \ldots, P_r$ forms partition of the element set, and the goal is to find a minimum cost sub-collection of sets such that atleast $k_i$ elements are covered from partition $P_i$. For that we can get a $H(\sum_{P_i} k_i)$-approximation by directly extending Slavík's [20] greedy approach, and unless P=NP we cannot do any better.